\documentclass[11pt, conference]{IEEEtran}
\usepackage{subfigure,stfloats}
\usepackage{epsfig,graphicx,psfrag}
\usepackage{amsfonts,amsmath,amssymb}
\usepackage{courier}

\newtheorem{theorem}{Theorem}[section]

\newtheorem{corollary}[theorem]{Corollary}

\begin{document}

\title{Using a Secret Key to Foil an Eavesdropper}

\author{
\authorblockN{Paul Cuff}
\authorblockA{Department of Electrical Engineering \\
Princeton University\\
E-mail: cuff@princeton.edu }
}

\maketitle

\begin{abstract}
This work addresses private communication with distributed systems in mind.  We consider how to best use secret key resources and communication to transmit signals across a system so that an eavesdropper is least capable to act on the signals.  One of the key assumptions is that the private signals are publicly available with a delay---in this case a delay of one.  We find that even if the source signal (information source) is memoryless, the design and performance of the optimal system has a strong dependence on which signals are assumed to be available to the eavesdropper with delay.

Specifically, we consider a distributed system with two components where information is known to only one component and communication resources are limited.  Instead of measuring secrecy by ``equivocation,'' we define a value function for the system, based on the actions of the system and the adversary, and characterize the optimal performance of the system, as measured by the average value obtained against the worst adversary.  The resulting optimal rate-payoff region is expressed with information theoretic inequalities, and the optimal communication methods are not standard source coding techniques but instead are methods that stem from synthesizing a memoryless channel.
\end{abstract}

\section{Introduction}
\label{section introduction}

Consider a situation where an adversary attempts to disrupt a distributed system.  The adversary may be attempting to jam communications, destabilize a power grid, or counter a miliary attack.  We investigate the nature of the communication used to control the system.  How should the system use secret key resources to establish coordinated behavior?  How should the system disseminate information about availability of power in a power grid, frequency channels or network routes used for communication, or military strategy adjustments?  This work establishes a new approach for defining secrecy in distributed systems, robust against an attacker who can causally observe the behavior of the system and apply statistical attacks to make use of the intercepted communication signals.  We find that the optimal way to communicate with limited encryption capability (secret key) is to reveal a distorted version of the information in the clear and use the encryption resources to hide the most important aspects of the information.

A large body of theoretical research provides a foundation for understanding privacy in communication.  The bulk of this research focusses on creating private channels from limited physical resources. A formal information theoretic study began with Shannon's 1949 paper, ``Communication Theory of Secrecy Systems'' \cite{ShannonSecrecy49}, where he established that the Vernam cipher using a one-time pad was necessary and sufficient for perfect secrecy.  Thus, we understand how to use a random secret key to create a private communication channel.  But how do we create and distribute this secret key?  Diffie and Hellman gave birth to modern cryptographic methods in 1976 \cite{Diffie76newdirections} by showing how to use ``trapdoor functions'' to enable key distribution using communication over public channels.  The resulting keys are protected by computational complexity but are not theoretically proven to be secret.  Nevertheless, they have withstood tests of time and enabled the broad commercial use of encryption.  Some different approaches for distributing secret keys use quantum channels \cite{BennettBrassard84} or observations of correlated variables \cite{gacs-korner73} \cite{csiszar-narayan00}.

Another method for creating private communication channels has had extensive recent interest.  Without the use of a shared secret key, channel noise can be used to hide a message if the noise is different for the eavesdropper than it is for the intended receiver.  This so-called ``physical layer security'' began with Wyner's famous wiretap channel \cite{WynerWiretap75}.  Now a variety of situations are studied, with multiple sources of information, fading, or multiple parties involved (see for example \cite{bloch08}).

In this work we do not worry about establishing a secret key or creating a private channel.  We assume that such a resource is already available, and we investigate how to best use it.  Just as channel coding and source coding share a complementary relationship---channel capacity characterizes the creation of communication resources from physical constraints, and rate-distortion theory establishes the best use of such resources---we see this work as a complementary development to the main body of information theoretic secrecy research, which is typically concerned with creating private communication resources.  We show how to use those resources to achieve optimal performance in a particular broad setting.

In our framework, the transmitter and intended receiver of communication are part of a distributed system.  A sequence of information is known at the transmit node, but the objective may not be to send that information to the receiver node verbatim.  Instead, the receiver will be producing a sequence of actions important to the distributed system that should be somehow correlated with the information sequence.  This is captured by a payoff function, the average value of which the communication system is designed to maximize.  So far this falls exactly in the category of rate-distortion theory.  However, the catch is that an adversary, who is also observing the communication over the public channel, is able to perform actions that will also affect the payoff.

This view of a secrecy system puts the adversary and the communication system in a zero-sum game against each other.  Indeed, when talking about security, a game theoretic formulation seems appropriate.  The optimal communication system in our framework is designed to maximize the average payoff against the most clever adversary.  Measurements of the uncertainty of the information with respect to an adversary, such as ``equivocation,'' do not indicate how useful the intercepted information is.  We work directly with an operational quantity---average payoff.

The main result, given in Section \ref{section main result}, corresponds to an encoding scheme that is quite different from optimal encoding techniques commonly used for source coding.  A reasonable question to explore is, ``which parts of problem statement are pivotal for these results?''  We find, for example, that the information we assume is available to the adversary is of primary importance.  In our framework, we assume the adversary not only intercepts the public communication but also observes the past information and actions of the system.  If we weaken the adversary by reducing the information that is available, then the results change substantially.  This is explored in Section \ref{section limited adversary}.

On the other hand, some details of the problem statement are not crucial.  Although our problem formulation defines the payoff of the system to be a function of three variables---the information, the action of the intended receiver, and the action of the adversary---a small modification would be to have two objectives to separately handle the behavior of the system and the adversary.  The first objective could be a distortion constraint at the intended receiver, and the second objective could be to force a level of distortion on the adversary.  In this case, the two distortion functions, each of only two variables, take the place of the payoff function.  It may seem that this special case would allow for a simplification of the main result given in Section \ref{theorem main result}, allowing for a much more basic optimal communication scheme.  Surprising, this does not seem to be true (see Section \ref{subsection no interaction}).

\section{Problem Statement}
\label{section problem statement}

\begin{figure}
\psfrag{a}[][][0.9]{$X^n$}
\psfrag{b}[][][0.9]{$Y^n$}
\psfrag{c}[][][0.9]{$Z^n$}
\psfrag{d}[][][0.9]{$J \in [2^{nR}]$}
\psfrag{e}[][][0.9]{$K \in [2^{nR_0}]$}
\psfrag{f}[][][0.8]{Node A}
\psfrag{g}[][][0.8]{Node B}
\psfrag{h}[][][0.7]{Adversary}
\centering
\includegraphics[width=.45\textwidth]{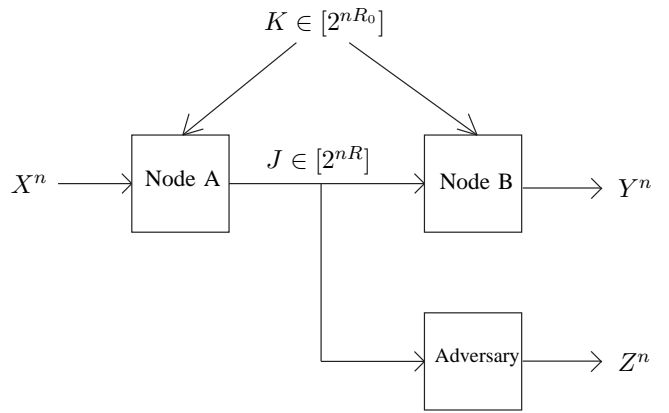}
\caption{{\em Distributed System with Adversary.}  The information sequence $X^n$ is i.i.d according to $p_0(x)$.  Node A and Node B are designed so that Node B can produce a sequence of actions $Y^n$ which depend on $X^n$.  The resources available to them are communication over a public channel at rate $R$ bits per action and secret key (independent of $X^n$) at rate $R_0$ bits per action.  An adversary taps into the message sent over the public channel and observes the actions of the system and information sequence causally.  The adversary attacks the system by minimizing a value function $\pi(x,y,z)$, and the system is designed to maximize the worst case average value obtained.}
\label{figure diagram}
\end{figure}

In Figure \ref{figure diagram}, a distributed system is represented by two components, Node A and Node B.  A third node, Node C, represents a hypothetical adversary who attacks the system.  An i.i.d. source of information $\{X_i\}_{i=1}^{\infty}$ is known to Node A.  The distribution of each element of the sequence $X_i$ is given by the probability mass function $p_0(x)$.

Node B and Node C take actions in the system.  A function $\pi(x,y,z)$ represents the value obtained by the system during each time epoch when the information symbol $X_i$ is $x$, the action $Y_i$ by the cooperating agent at Node B is $y$, and the attack $Z_i$ by the adversary is $z$.  The system, which operates over blocks of length $n$, is designed to maximize the average value
\begin{eqnarray*}
\overline{\Pi} & = & \mathbf{E} \; \frac{1}{n} \sum_{i=1}^n \pi (X_i,Y_i,Z_i),
\end{eqnarray*}
while the adversary tries to minimize $\overline{\Pi}$.  One interpretation of $\pi$ is as the payoff function of a zero-sum game between the distributed system and the adversary.

The communication in the system is defined as follows.  Node A sends a message to help Node B coordinate actions with the source sequence $\{X_i\}_{i=1}^{\infty}$, but the message is available to both agents, Node B and Node C.  However, a secret key (independent of the source sequence) is known only to Nodes A and B, which can be used to establish secrecy and coordination in the system.  For example, if the secret key is large enough, the communication can be fully encrypted so that only Node B can make use of it.

An $(R_0, R, n)$ coordination scheme consists of an encoder and a decoder utilizing a secret key rate of $R_0$ bits per symbol and a description rate of $R$ bits per symbol.  The encoder at Node A transmits an $nR$-bit message $J \in [2^{nR}]$ based on the source realization $X^n$ and an $nR_0$-bit secret key $K \in [2^{nR_0}]$ which is independent of the source.  The encoding at Node A can be designed to use randomization; thus, it is described by a conditional probability function $p(j|x^n,k)$.

The attack by the adversary (Node C) on the distributed system (Nodes A and B) occurs interactively.  Nodes B and C first receive the communication produced by Node A.  Then they each produce one action, $Y_1$ and $Z_1$.  This constitutes the first instance of the game, after which both nodes are aware of the action taken by the other node and the first symbol of the source realization $X_1$.  At this point they each choose a second action, $Y_2$ and $Z_2$, and proceed in a similar manner.  For each iteration of the game, the decoder at Node B generates an action $Y_k$ based on the message $J$, the secret key $K$, and the past actions $X^{i-1}$, $Y^{i-1}$, and $Z^{i-1}$.  This decoder is described by a set of conditional probability distributions $\{p(y_i|j,k,x^{i-1},y^{i-1},z^{i-1})\}_{i=1}^{n}$.  The adversary (Node C) also generates actions in a similar way as the coordinating agent (Node B), except that he doesn't have access to the secret key.  His actions are described by a set of conditional probability distributions $\{p(z_i|j,x^{i-1},y^{i-1},z^{i-1})\}_{i=1}^{n}$.  We consider the strategy of the eavesdropper that inflicts the most damage on the system.  An $(R_0, R, n)$ coordination scheme is evaluated by the expected average payoff it assures against the worst-case adversary.

To summarize:
\begin{itemize}
\item Source:  $\{X_i\}$ i.i.d. $\sim p_0(x)$
\item $(R_0, R, n)$ Coordination Scheme:
\begin{itemize}
\item Key:  $K \sim Unif[2^{nR_0}]$ independent of source
\item Message:  $J \in [2^{nR}]$
\item Encoder (Node A):
\begin{eqnarray*}
C_A & \triangleq & p(j|x^n,k)
\end{eqnarray*}
\item Decoder (Node B):
\begin{eqnarray*}
C_B & \triangleq & \{p(y_i|j,k,x^{i-1},y^{i-1},z^{i-1})\}_{i=1}^{n}
\end{eqnarray*}
\end{itemize}
\item Adversary
\begin{itemize}
\item Strategy (Node C):
\begin{eqnarray*}
C_C & \triangleq & \{p(z_i|j,x^{i-1},y^{i-1},z^{i-1})\}_{i=1}^{n}
\end{eqnarray*}
\end{itemize}
\item Joint Distribution:  product of all of the above.
\item Average Value:
\begin{eqnarray*}
\overline{\Pi}_{p_0}(C_A,C_B,C_C) & \triangleq & \mathbf{E} \; \frac{1}{n} \sum_{i=1}^n \pi (X_i,Y_i,Z_i),
\end{eqnarray*}
\item Robust achievable value:
\begin{eqnarray*}
\Pi_{p_0}(R_0,R) & \triangleq & \sup_{\{n, C_A, C_B\}} \min_{C_C} \; \overline{\Pi}
\end{eqnarray*}
\end{itemize}

{\em Note:}
A system may not be aware of an adversary or the actions an adversary has taken.  There may be multiple adversaries attacking a system.  This problem statement defines a decoder at Node B that responds to a single adversary.  However, we will find that the optimal max-min codec does not take the actions of adversary into account, nor does it use the causal source information.  Decoders at Node B of the form $C_B = \{p(y_i|j,k)\}$ achieve optimality.  Therefore, these results are more widely applicable than the specific setting described.  They may also apply to situations where an adversary is not easily detected or multiple adversaries exist.

On the other hand, the causal information available to the adversary is crucial for these results.  If the adversary had less information available, the system can achieve the same value using less resources.  In the extreme case, if the adversary has no causal information of the actions at Node B or the information at Node A, then it is easy to transmit a message that is useless to the adversary.  Any small rate of secret key is as good as perfect secrecy in the sense that the adversary cannot mount an attack based on the intercepted message.  This is discussed further in Section \ref{section limited adversary}.

\section{Main Result}
\label{section main result}

\begin{theorem}
\label{theorem main result}
\begin{eqnarray*}
\Pi_{p_0}(R_0,R) & = & \max_{p(y,u,v|x) \in {\cal P}} \; \min_{z(u)} \mathbf{E} \; \pi(X,Y,z(U)),
\end{eqnarray*}
where
\begin{eqnarray*}
{\cal P}_{p_0}(R_0,R) & \triangleq & \left\{
\begin{array}{rcl}
p(y,u,v|x) & : & \\
p(y|u,v,x) & = & p(y|u,v), \\
R_0 & \geq & I(X,Y;V|U), \\
R & \geq & I(X;U,V). \\
\end{array}
\right\}.
\end{eqnarray*}
\end{theorem}

\section{Converse}
\label{section converse}

\begin{proof}
In this section we prove an upper bound on $\Pi_{p_0}(R_0,R)$.  For any coordination scheme satisfying the rate constraints $R_0$ and $R$, we identify random variables $X$, $Y$, $U$, and $V$ such that $p(y,u,v|x) \in {\cal P}_{p_0}(R_0,R)$ and
\begin{eqnarray*}
\min_{C_C} \overline{\Pi} & = & \min_{z(u)} \mathbf{E} \; \pi(X,Y,z(U).
\end{eqnarray*}

We first identify the random variables for the converse.  Let $Q$ be a random variable uniformly distributed on the set $[n]$ and independent of $(X^n,Y^n,Z^n,J,K)$.  We will use $Q$ as a random index for sequences, where $X_Q$ is a function of the sequence $X^n$ and the variable $Q$ that selects the $Q$th element of the sequence.  Notice that for an i.i.d. sequence like $X^n$, the random index $Q$ is independent of $X_Q$.
\\

\noindent
{\em Identification of variables:}
\begin{itemize}
\item $X = X_Q$
\item $Y = Y_Q$
\item $Z = Z_Q$
\item $V = K$
\item $U = \{J, X^{Q-1}, Y^{Q-1}, Z^{Q-1},Q\}$
\end{itemize}

\noindent
{\em Important properties:}
\begin{itemize}
\item Independence
\begin{eqnarray*}
X_Q & \perp & Q,
\end{eqnarray*}
\item Markovity
\begin{equation*}
(X_i^n, Z_i) \;\; - \;\; (J,K,X^{i-1},Y^{i-1},Z^{i-1}) \;\; - \;\; Y_i,
\end{equation*}
\begin{equation*}
(X_i^n, Y_i, K) \;\; - \;\; (J,X^{i-1},Y^{i-1},Z^{i-1}) \;\; - \;\; Z_i,
\end{equation*}
\begin{equation*}
(X_i) \;\; - \;\; (J,K,X^{i-1}) \;\; - \;\; (Y^i,Z^i).
\end{equation*}
\end{itemize}

We start by noticing that $X$ is distributed according to $p_0$ and $X - (U,V) - Y$ form a Markov chain, according to the definitions and properties above.

Next, we show the inequalities that involve $R_0$ and $R$.

\begin{eqnarray*}
n R_0 & \geq & H(K) \\
& \geq & H(K|J) \\
& \geq & I(X^n,Y^n,Z^n;K|J) \\
& = & \sum_{q=1}^n I(X_q,Y_q,Z_q;K|J,X^{q-1},Y^{q-1},Z^{q-1}) \\
& = & n \; I(X_Q,Y_Q,Z_Q;K|J,X^{Q-1},Y^{Q-1},Z^{Q-1},Q) \\
& = & n \; I(X_Q,Y_Q;K|J,X^{Q-1},Y^{Q-1},Z^{Q-1},Q) \\
& = & n \; I(X,Y;V|U).
\end{eqnarray*}
\begin{eqnarray*}
n R & \geq & H(J) \\
& \geq & H(J|K) \\
& \geq & I(X^n;J|K) \\
& = & I(X^n;J,K) \\
& = & \sum_{q=1}^n I(X_k;J,K,X^{k-1}) \\
& = & \sum_{q=1}^n I(X_k;J,K,X^{k-1},Y^{k-1},Z^{k-1}) \\
& = & n \; I(X_Q;J,K,X^{Q-1},Y^{Q-1},Z^{Q-1},Q) \\
& = & n \; I(X;U,V).
\end{eqnarray*}

Consequently, the variables $U$, $V$, and $Y$ are conditionally distributed according to $p(y,u,v|x) \in {\cal P}_{p_0}(R_0,R)$.  Now consider that
\begin{eqnarray*}
\min_{C_C} \overline{\Pi} & = & \min_{\{p(z_i|j,x^{i-1},y^{i-1},z^{i-1})\}_{i=1}^n} \mathbf{E} \; \frac{1}{n} \sum_{i=1}^n \pi (X_i,Y_i,Z_i) \\
& = & \min_{p(z_q|j,x^{q-1},y^{q-1},z^{q-1},q)} \mathbf{E} \; \mathbf{E} \; \left[ \pi(X_Q,Y_Q,Z_Q) | Q \right] \\
& = & \min_{p(z|u)} \mathbf{E} \; \pi(X,Y,Z) \\
& = & \min_{z(u)} \mathbf{E} \; \pi(X,Y,z(U)).
\end{eqnarray*}
\end{proof}

\section{Sketch of Achievability}
\label{section achievability}

\begin{proof}
Here we design a system that guarantees an average expected reward approaching the value of $\Pi_{p_0}(R_0,R)$ given in Theorem \ref{theorem main result}.  This is done using the notions of empirical coordination and strong coordination discussed in \cite{cuff-permuter-cover10}.

Begin with the optimal conditional distribution $p(y,u,v|x) \in {\cal P}_{p_0}(R_0,R)$.  The main idea is to first specify a $U^n$ sequence that is empirically coordinated with $X^n$, which is to say that it is jointly typical with high probability, using a communication rate of roughly $I(X;U)$ bits per source symbol.  Then produce a sequence $Y^n$ that is strongly coordinated with $X^n$, conditioned on $U^n$, which is to say that $X^n$ and $Y^n$ appear to be memoryless, even with full knowledge of the codebook used for coordination.  The variable $V$ is an auxiliary variable that only has meaning in the process of achieving strong coordination.  The rates needed for strong coordination over a public channel are $I(X,Y;V|U)$ bits of secret key per symbol and $I(X;V|U)$ bits of communication per symbol, with the condition that $X - (U,V) - Y$ form a Markov chain.  These rate requirements are touched on in \cite{cuff08}.

After encoding, the adversary knows the sequence $U^n$.  The other sequences in the system, $X^n$ and $Y^n$, are correlated with $U^n$, but otherwise appear to be nearly memoryless (in total variation), even conditioned on everything known by the adversary.  Therefore, the best strategy for the adversary to minimize the average value to the system will not be substantially better than choosing the best strategy $z(u)$ that minimizes $\mathbf{E} \; \pi (X,Y,z(U))$ and applying this strategy during each iteration.
\end{proof}

\section{Special Cases}
\label{section special cases}

\subsection{Lossless}
\label{subsection lossless}

In some cases of $\pi$, the description of $\Pi_{p_0}(R_0,R)$ simplifies.  A particular important case is the lossless setting, where $Y^n$ is required to be equal to $X^n$ with high probability.\footnote{This would be the case if $\pi(x,y,z)$ took on very large negative values for $x \neq y$.}  This is worth considering because it resembles the familiar setting usually considered for information theoretic secrecy, where the information source sequence $X^n$ must be recovered by the intended receiver.

In the lossless case, $\pi$ can be though of as a distortion function that is being inflicted on an eavesdropper.  The system is designed to maximize the distortion.

The following corollary is found in \cite{cuff10} and can be reduced to a linear program.

\begin{corollary}
\label{corollary lossless}
For the lossless case, where $Y^n$ must equal $X^n$ with high probability, the robust average value is
\begin{eqnarray*}
\Pi_{p_0}(R_0,R) & = & \max_{p(u|x) \in {\cal P}} \min_{z(u)} \mathbf{E} \; \pi(X,z(U)),
\end{eqnarray*}
where
\begin{eqnarray*}
{\cal P}_{p_0}(R_0,R) & \triangleq & \left\{
\begin{array}{rcl}
p(u|x) & : & \\
R_0 & \geq & H(X|U), \\
R & \geq & H(X). \\
\end{array}
\right\}.
\end{eqnarray*}
\end{corollary}

\subsection{No interaction}
\label{subsection no interaction}

We can imagine some cases where the intended receiver and the adversary each attempt to reconstruct the source $X^n$ with low distortion, and they are each not concerned with the reconstruction of the other.  This situation can be expressed with two separate value functions $\pi_1(x,y)$ and $\pi_2(x,z)$.  We might ask for two separate constraints to be satisfied or simply construct a value function $\pi(x,y,z)$ that is the product or sum of these two separate components.

By removing the direct interaction between the intended receiver and the adversary, it might seem that a simpler encoding scheme, not involving strong coordination, is optimal.  Unfortunately, that is not the case.  If the distortion constraint of the intended receiver were achieved in a careless way, without using strong coordination, then the adversary would be able to infer extra information about $X^n$ indirectly through observation of the past actions of the intended receiver.  This would help the adversary to reduce its own distortion.

\section{Limited Adversary}
\label{section limited adversary}

The result expressed in Theorem \ref{theorem main result} specifies a guaranteed average value against an adversary who has causal information of all actions and information in the system.  The results change significantly if the adversary is limited in the information available.  Here we give three results related to the main result in Theorem \ref{theorem main result}, with proofs omitted.

\begin{theorem}
\label{theorem past y}
Consider an adversary who only has access to the message $J$ and the past actions at Node B, but not the past information symbols.  Thus, the strategies of the adversary are defined by $\{p(z_i|j,y^{i-1},z^{i-1})\}_{i=1}^n$.
\begin{eqnarray*}
\Pi_{p_0}(R_0,R) & = & \max_{p(y,u,v|x) \in {\cal P}} \min_{z(u)} \mathbf{E} \; \pi(X,Y,z(U)),
\end{eqnarray*}
where
\begin{eqnarray*}
{\cal P}_{p_0}(R_0,R) & \triangleq & \left\{
\begin{array}{rcl}
p(y,u,v|x) & : & \\
p(y|u,v,x) & = & p(y|u,v), \\
R_0 & \geq & I(Y;V|U), \\
R & \geq & I(X;U,V). \\
\end{array}
\right\}.
\end{eqnarray*}
\end{theorem}

\begin{theorem}
\label{theorem past x}
Consider an adversary who only has access to the message $J$ and the past information symbols, but not the past actions at Node B.  Thus, the strategies of the adversary are defined by $\{p(z_i|j,x^{i-1},z^{i-1})\}_{i=1}^n$.
\begin{eqnarray*}
\Pi_{p_0}(R_0,R) & = & \max_{p(y,u|x) \in {\cal P}} \min_{z(u)} \mathbf{E} \; \pi(X,Y,z(U)),
\end{eqnarray*}
where
\begin{eqnarray*}
{\cal P}_{p_0}(R_0,R) & \triangleq & \left\{
\begin{array}{rcl}
p(y,u|x) & : & \\
R_0 & \geq & I(X,Y|U), \\
R & \geq & I(X;U,Y). \\
\end{array}
\right\}.
\end{eqnarray*}
\end{theorem}

\begin{theorem}
\label{theorem no info}
Consider an adversary who only has access to the message $J$.  Thus, the strategies of the adversary are defined by $\{p(z_i|j,z^{i-1})\}_{i=1}^n$.
\begin{eqnarray*}
\Pi_{p_0}(R_0,R) & = & \max_{p(y|x) \in {\cal P}} \min_{z} \mathbf{E} \; \pi(X,Y,z),
\end{eqnarray*}
where
\begin{eqnarray*}
{\cal P}_{p_0}(R_0,R) & \triangleq & \left\{
\begin{array}{rcl}
p(y|x) & : & \\
R_0 & > & 0, \\
R & \geq & I(X;Y). \\
\end{array}
\right\}.
\end{eqnarray*}
\end{theorem}

The third setting, in which the adversary has no causal information, is the approach taken by Yamamoto in \cite{YamamotoCipherDistortion97}.  We see from Theorem \ref{theorem no info} that there is no non-trivial lower bound on the rate of secret key needed.  In Theorem 3 of \cite{YamamotoCipherDistortion97} a lower bound is provided, but after careful consideration and with the proper choice of auxiliary random variables, that bound can be shown to be zero.

In other work by Yamamoto \cite{yamamoto88}, $R_0$ is taken to be exactly zero.  The results do not have the form of Theorem \ref{theorem no info}.  Instead they coincide with Theorems \ref{theorem main result}, \ref{theorem past y}, and \ref{theorem past x} with $R_0 = 0$.  This illustrates a discontinuity of $\Pi_{p_0}(R_0,R)$ at $R_0 = 0$.

\section{Summary}
\label{section summary}

We've identified the optimal use of communication and secret key resources for a distributed system to compete in what amounts to a zero-sum game with an adversary.  This brings new insight into the theoretical limits of secrecy systems.  The results found here are not directly related to information measures such as ``equivocation,'' which are commonly used in the literature.

Perhaps the most surprising, yet subtle consequence of the main result is that the combined resources of public communication and secret key of equal rates is strictly superior to a single {\em private} channel resource of the same rate.  Furthermore, a secret key rate in excess of the public communication rate can actually be useful.

\bibliographystyle{IEEEtran}
\bibliography{ref}

\end{document}